\documentclass[conference]{IEEEtran.1.8b}
\usepackage{listings}
\usepackage[colorlinks,linkcolor=blue,citecolor=blue,urlcolor=blue]{hyperref}

\usepackage{color}
\definecolor{myblue}{rgb}{0.2,0.4,0.8}
\definecolor{mygreen}{rgb}{0.3,0.7,0.3}
\definecolor{myred}{rgb}{0.8,0.4,0.2}
\definecolor{myyellow}{rgb}{0.7,0.7,0.3}
\definecolor{mygray}{rgb}{0.6,0.6,0.6}

\usepackage{ifpdf}

\usepackage{cite}

\ifCLASSINFOpdf
  \usepackage[pdftex]{graphicx}
  \graphicspath{{images/}}
\else
  \usepackage[dvips]{graphicx}
\fi
\usepackage{algpseudocode}
\usepackage{algorithmicx}

\usepackage{tikz}
\usetikzlibrary{arrows,shapes.geometric,shapes.symbols}

\tikzset{
	start-end/.style={
		draw,
		rectangle,
		rounded corners,
	},
	input/.style={ 
		draw,
		trapezium,
		trapezium left angle=60,
		trapezium right angle=120,
	},
	operation/.style={
		draw,
		rectangle
	},
	loop/.style={ 
		draw,
		chamfered rectangle,
		chamfered rectangle xsep=2cm
	},
	decision/.style={ 
		draw,
		diamond,
		aspect=#1
	},
	decision/.default=1,
	print/.style={ 
		draw,
		tape,
		tape bend top=none
	},
	connection/.style={
		draw,
		circle,
		radius=5pt,
	},
	process rectangle outer width/.initial=0.15cm,
	predefined process/.style={
		rectangle,
		draw,
		append after command={
			\pgfextra{
				\draw
				($(\tikzlastnode.north west)-(0,0.5\pgflinewidth)$)--
				($(\tikzlastnode.north west)-(\pgfkeysvalueof{/tikz/process rectangle outer width},0.5\pgflinewidth)$)--
				($(\tikzlastnode.south west)+(-\pgfkeysvalueof{/tikz/process rectangle outer width},+0.5\pgflinewidth)$)--
				($(\tikzlastnode.south west)+(0,0.5\pgflinewidth)$);
				\draw
				($(\tikzlastnode.north east)-(0,0.5\pgflinewidth)$)--
				($(\tikzlastnode.north east)+(\pgfkeysvalueof{/tikz/process rectangle outer width},-0.5\pgflinewidth)$)--
				($(\tikzlastnode.south east)+(\pgfkeysvalueof{/tikz/process rectangle outer width},0.5\pgflinewidth)$)--
				($(\tikzlastnode.south east)+(0,0.5\pgflinewidth)$);
			}  
		},
		text width=#1,
		align=center
	},
	predefined process/.default=1.75cm,
	man op/.style={ 
		draw,
		trapezium,
		shape border rotate=180,
		text width=2cm,
		align=center,
	},
	extract/.style={
		draw,
		isosceles triangle,
		isosceles triangle apex angle=60,
		shape border rotate=90
	},
	merge/.style={
		draw,
		isosceles triangle,
		isosceles triangle apex angle=60,
		shape border rotate=-90
	},
}

\usepackage{array,tabularx}

\newcommand{\myfig}[5]
{
\begin{figure}[t]
\begin{center}
\ifpdf
\includegraphics[width=#4\linewidth]{#1}
\else
\includegraphics[width=#4\linewidth]{#1}
\fi
\end{center}
\vspace{#5}
\caption{#2}\label{#3}
\end{figure}
}

\newcolumntype{R}[1]{>{\raggedleft\arraybackslash}m{#1}}

\newcommand{\scbf}[1]{\vspace {0.05in}\noindent{\textbf{#1.}}}

\hypersetup{draft}

\begin{document}
%
\title{Talos: Neutralizing Vulnerabilities with Security Workarounds for Rapid Response}


\author{
\IEEEauthorblockN{Zhen Huang\hspace{0.25in} 
	Mariana D'Angelo\hspace{0.25in}
	Dhaval Miyani\hspace{0.25in}
	David Lie}
\IEEEauthorblockA{University of Toronto\\
	\textit{\{z.huang,mariana.dangelo,dhaval.miyani\}@mail.utoronto.ca,lie@eecg.toronto.edu}}
}


%


\maketitle

\begin{abstract}
There is often a considerable delay between the discovery of a vulnerability and the issue of a patch.  One way to mitigate this window of vulnerability is to use a configuration workaround, which prevents the vulnerable code from being executed at the cost of some lost functionality -- but {\em only} if one is available.  Since application configurations are not specifically designed to mitigate software vulnerabilities, we find that they only cover 25.2\% of vulnerabilities.


To minimize patch delay vulnerabilities and address the limitations of configuration workarounds, we propose Security Workarounds for Rapid Response (SWRRs), which are designed to neutralize security vulnerabilities in a timely, secure, and unobtrusive manner. Similar to configuration workarounds, SWRRs neutralize vulnerabilities by preventing vulnerable code from being executed at the cost of some lost functionality. However, the key difference is that SWRRs use existing error-handling code within applications, which enables them to be mechanically inserted with minimal knowledge of the application and minimal developer effort.  This allows SWRRs to achieve high coverage while still being fast and easy to deploy.

We have designed and implemented Talos, a system that mechanically instruments SWRRs into a given application, and evaluate it on five popular Linux server applications.  We run exploits against 11 real-world software vulnerabilities and show that SWRRs neutralize the vulnerabilities in all cases.  Quantitative measurements on 320 SWRRs indicate that SWRRs instrumented by Talos can neutralize 75.1\% of all potential vulnerabilities and incur a loss of functionality similar to configuration workarounds in 71.3\% of those cases.  Our overall conclusion is that automatically generated SWRRs can safely mitigate 2.1$\times$ more vulnerabilities, while only incurring a loss of functionality comparable to that of traditional configuration workarounds. 

\end{abstract}


%

\section{Introduction}
Patches are the commonly-accepted solution for completely preventing a security vulnerability from being exploited.  Patches fix security vulnerabilities and, in most cases, do so with no loss of functionality or performance for the application.  However, despite their benefits, patches are not perfect.

An often ignored drawback of patches is the {\em pre-patch window of vulnerability} that exists between the time a vulnerability is discovered and the time a patch is issued. This vulnerability window is inherent to security patches because patches must be manually created and tested, which takes time and effort to do correctly.  Although a large number of techniques have been proposed to automatically generate patches to fix vulnerabilities~\cite{Perkins2009,Weimer2009, Wei2010, Goues2012-genprog, Goues2012-study, Nguyen2013, Shahriar2012}, to the best of our knowledge, none have been widely adopted in practice.  As a recent study~\cite{Shahzad2012} and our own findings in Section~\ref{sec:study} demonstrate, the length of this window can be significant, and is unlikely to decrease on average due to the complexity of creating a security patch.  While the risk of exploitation during the pre-patch window can be reduced by keeping the vulnerability secret, this is just security-through-obscurity.  As the highly active market of zero-day vulnerabilities demonstrates, there is no shortage of instances where attackers may be aware of and be able to exploit vulnerabilities during this window~\cite{Bilge2012}.

While the period before a vulnerability is known can be reduced by better vulnerability detection and software engineering practices, we believe we can address the window of vulnerability that exists between the time the vulnerability is known to the developer (or to the public) and when the patch is issued.  To do this, we take inspiration from configuration workarounds, which are a commonly used mechanism to address the pre-patch window of vulnerability.  Configuration workarounds disable functionality related to the vulnerable code so that it cannot be triggered by an attacker.  For example, the Android Stagefright bug, discovered in 2015, is a perfect example of this.  This remote code execution vulnerability, which affected almost 1 billion devices, was discovered as early as April 2015, although not publicly disclosed until July.  A patch was not available until August 2015, several months after the vulnerability was discovered, and even at the time of writing, many smartphones models still do not have a usable patch~\cite{Stagefright1, Stagefright2}.   Fortunately, the worst methods of exploitation via a malicious MMS can be prevented by configuring the MMS client on the phone to not automatically download media files from MMS messages.  In exchange for some minor loss of functionality, the configuration workaround protects the user from the exploitation of a very serious vulnerability.


However, configuration workarounds are far from ideal for mitigating security vulnerabilities.  Again, as we show in Section~\ref{sec:study}, configuration workarounds have low coverage -- there are many vulnerabilities for which no configuration workaround exists because there is no configuration option that can disable the vulnerability.  Configuration options are designed to allow users to easily alter the behaviour of a program, and thus only cover functionality that most users would like to enable or disable.  Thus it is hardly surprising that very few vulnerabilities have configuration workarounds, and it essentially becomes a matter of random serendipity whether a vulnerability can be neutralized with a configuration workaround or not.

Motivated by the problems of the pre-patch vulnerability window and the low coverage of configuration workarounds, we propose {\em Security Workarounds for Rapid Response (SWRRs)}, workarounds that can be mechanically generated to address a large percentage of vulnerabilities.  The main challenge in designing SWRRs is that they must work in a broad range of circumstances.  By their nature, vulnerabilities are not known a priori, and thus SWRRs must work for any vulnerability that can occur.  Another challenge is that vulnerabilities can occur anywhere in an application, and be related to almost any type of functionality, but an SWRR must ensure that the application continues to work after the SWRR is applied.  Thus, we design SWRRs to be simple and generic, relying on very few assumptions about either applications or vulnerabilities.  The cost of this generality is that, like configuration workarounds, users must be willing to accept some minor loss of functionality in return for protection from a vulnerability until a patch is issued.  

Our key insight for making SWRRs generic and cheap to deploy is that application error-handling code, whose purpose is to gracefully return an application to a good state when an unexpected error arises, can be found and invoked using static analysis.  Based on this insight, we have designed and implemented a system called Talos, which detects such error-handling code using static analysis and adds SWRRs into a given application. Each SWRR prevents the execution of code where a vulnerability is located, and calls the error-handling code instead.  With Talos, developers can deploy SWRRs either as patches or in-place as part of an application so that they can be activated with run-time loadable configurations.

In summary, SWRRs provide benefits for both software developers and users at a small cost. For the cost of having to run Talos and issue the resulting patch, software developers benefit by having a solution that protects their users; this affords them more time and less immediate pressure to create, test, and deploy a patch.  Users benefit by having a solution that protects them during the pre-patch vulnerability window at the cost of having to accept some loss of functionality.  In addition, in cases where users cannot install a patch for compatibility reasons or where no patch exists because the software is no longer supported, users can still use an SWRR to protect themselves.

\noindent This paper makes the following contributions:
\begin{enumerate}
	\item We propose SWRRs, which provide a low-cost way for software developers to quickly protect users during the pre-patch vulnerability window.
	\item We design and implement a software tool called Talos to demonstrate that SWRRs can be practically deployed.  To safely continue the execution of an application, Talos heuristically identifies error-handling code in the program and transfers execution to those paths to avoid having to execute vulnerable code.
	\item We evaluate the effectiveness and coverage of the SWRRs inserted by Talos into 5 popular applications.  When tested against 11 vulnerabilities, SWRRs generated by Talos successfully neutralize the vulnerabilities in all cases.  Empirical tests on 320 Talos-generated SWRRs show that they can achieve an effective coverage that is 2.1$\times$ that of traditional configuration workarounds.	
\end{enumerate}

We begin in Section~\ref{sec:study} with a study based on data we collected that demonstrates the motivation behind SWRRs.  Then we give an overview of SWRRs in Section~\ref{sec:overview} and describe Talos, our tool for automatically inserting SWRRs into application source code, in Section~\ref{sec:talos}. We provide details about the implementation of Talos in Section~\ref{sec:implementation}.  We then evaluate the SWRRs that Talos instruments into applications in Section~\ref{sec:evaluation}.  We discuss the limitations and other issues of SWRRs in Section~\ref{sec:discussion}.  We then provide a comparison with related work in Section~\ref{sec:related} and conclude in Section~\ref{sec:conclusion}.

\section{Motivation}
\label{sec:study}
\subsection{The pre-patch vulnerability window}\label{sec:patch_study}

\begin{table}
	\begin{center}
		\caption{Security Patch Statistics.}
		\begin{tabular}{|l|r|r|r|r|r|}
			\hline
			\textbf{App.} & \textbf{Vulns.} & \textbf{Delay (Days)} & \textbf{SLOC} & \textbf{Funcs} & \textbf{Files} \\
			\hline
			lighttpd & 27 & 54 & 49 & 2 & 2 \\
			\hline
			apache & 30 & 61 & 47 & 2 & 2 \\
			\hline
			squid & 46 & 73 & 64 & 6 & 3 \\
			\hline
			proftpd & 16 & 9 & 95 & 4 & 2 \\
			\hline
			sqlite & 12 & 62 & 17 & 4 & 3 \\
			\hline
			\textbf{AVERAGE} & 26 & 52 & 54 & 4 & 2 \\
			\hline
		\end{tabular}
		
		\label{tbl:patch_study}
	\end{center}
\end{table}

We begin with a study of the lifecycle and complexity of software patches for recent security vulnerabilities. The vulnerabilities used in our study were collected from various sources, including common vulnerability databases~\cite{NVD, MITRE, CVEdetails, securitytracker}, vendor-specific security bulletins~\cite{MSSecBulletin, OracleSecAlerts, ApacheHTTPDSec}, and software bug databases~\cite{ProftpdBugs, RedHatBugzilla, DebianBugs}. 

For our study, we need information on the disclosure date of vulnerabilities, the release/commit date of patches, and the source code of the patches. Hence we choose open-source applications that are popular, reasonably complex, mature, being actively developed and maintained, and have a decent number of vulnerabilities. For each application, we selected as many vulnerabilities as possible that have the required information for manual examination. Our results are shown in Table~\ref{tbl:patch_study}. Column ``Vulns.'' shows the number of examined vulnerabilities. Column ``Delay'' shows the average number of days between the disclosure of security vulnerabilities and the release of corresponding software patches.   We obtain the date when a vulnerability is disclosed from either an official vulnerability disclosure or the bug report for the vulnerability.  For some vulnerabilities, we could not find an official disclosure date, so we approximate this using the earliest dated document in which they are referenced.  From the collected data, we can see that it takes considerable time to release a patch and the size of the pre-patch vulnerability window is significant, averaging around 1.5 months after the vulnerability is disclosed. 

We find that 43.4\% of the vulnerabilities were patched within 7 days after the vulnerabilities were disclosed, 23.3\% of them were patched between 7 days and 30 days, and 33.3\% of them were patched after 30 days. Similarly, a recent study on the lifecycle of security vulnerabilities in operating systems and web browsers shows that among open source vendors, 65\% of the vulnerabilities were patched within 7 days, 9\% of them were patched between 7 days and 30 days, and 18\% of them were patched after 30 days~\cite{Shahzad2012}. Both our study and their study indicate that a significant number of vulnerabilities were patched after 30 days.

To understand the bottleneck in releasing a patch, we break the task of releasing a patch into steps including \textit{vulnerability triage}, \textit{constructing a patch}, and \textit{regression test}. We study the time spent in each step by examining the bug reports of vulnerabilities. Unfortunately, we were only able to locate the bug reports for 21 of the vulnerabilities that are shown in Table~\ref{tbl:patch_study}. 

As most of these bug reports do not contain the time when a developer was assigned or when a tester was assigned, we conservatively assign the period of time between when the bug is reported and when the first patch attempt is created as the time spent for vulnerability triage, the period of time between when the first patch attempt is created and when the last patch attempt is created as the time spent to construct a patch, and the period of time between when the last patch attempt is created and when the patch is committed as the time spent for regression test. 

For these vulnerabilities, we find that the time and effort spent in constructing a patch is very significant. For the 8 vulnerabilities that took more than one day to create a patch, 89\% of the time was spent in constructing a patch. And 9 of the vulnerabilities took between two to six attempts to patch correctly.
Particularly the bug report of one proftpd vulnerability (CVE-2012-6095 \cite{CVE-2012-6095}) contains five patch attempts, of which the last patch attempt was created 96 days after the first patch attempt was created, and 29 comments from the developer and the testers, of which the comments from the developer along the time line include ``\textit{This updates the previous patch ...}'', ``\textit{This patch builds on the previous one ...}'', ``\textit{I've just committed more changes ...}'', ``\textit{Hopefully the combination of ... addresses the remaining issues.}'', ``\textit{Unfortunately I don't have a good/easy fix/solution for this yet.}'', and one of the very last comments from the testers is still ``\textit{I'm afraid I found a bug in ...}''. 

To understand further why constructing a patch is non-trivial, we further study the complexity of the patches, which are available for all the vulnerabilities shown in Table~\ref{tbl:patch_study}. We use column ``SLOC'' to show the number of lines of source code in patches, column ``Funcs'' to show the number of functions that are changed by patches, and column ``Files'' to show the number of source code files that are changed by patches. We find that on average patches consist of 54 lines of source code and span 4 functions in 2 source code files. This suggests that on average, patches involve non-trivial changes to the application code.  As a result, the vulnerability window is likely inherent to patches, as time must be spent by human engineers to understand the vulnerability, design and implement the fix, and finally test and review the patch before release.  Due to our need for detailed bug reports and source code patches in performing this study, we were restricted to open-source applications.  However, we found no evidence that these conclusions are restricted to open-source projects, and so we believe they should apply equally to both open- and closed-source applications.

\subsection{Configuration workaround coverage of vulnerabilities}

\begin{table}
\begin{center}
	\caption{Configuration workaround statistics.}
\begin{tabular}{|l|r|r|r|l|}
\hline
\textbf{App.} & \textbf{Options} & \textbf {Vulns.} & \textbf{Workaround} & \textbf{Period} \\
\hline
lighttpd& 88 & 27 & 14.8\% & 2005-14 \\
\hline
apache & 74 & 42 & 16.7\% & 2002-14 \\
\hline
squid & 174 & 30 & 6.7\% & 2001-15 \\
\hline
proftpd & 28 & 20 & 20.0\% & 2004-13 \\
\hline
IE & 33 & 31 & 54.8\% & 2000-14 \\
\hline
Office & 325 & 32 & 37.5\% & 2000-11 \\
\hline
\end{tabular}
\label{tbl:study}
\end{center}
\end{table}

Since configuration workarounds represent the current best solution for mitigating the vulnerability window, we also present our study of configuration workarounds for recent security vulnerabilities. We define a configuration workaround as any vulnerability mitigation that involves modifying the configuration of the application (i.e., configuration options supported by the application) and exclude many other common fixes such as patching the application binary, disabling the application, or placing the vulnerability out of the reach of attackers (e.g., tightening firewall rules). 

For this study, we add two popular closed-source applications: Internet Explorer and Microsoft Office. We also exclude sqlite because sqlite does not support any configuration options. For each application, we again randomly select a number of vulnerabilities and search both the software vendors' websites and Internet to determine if a configuration workaround is available.  We tabulate the percentage of vulnerabilities examined for which we were able to find a configuration workaround as well as the time period over which the manually examined vulnerabilities were reported.  We also tabulate the number of configuration options for each application.  Table~\ref{tbl:study} presents the results. Column ``Options'' shows the number of configuration options that each application has. Column ``Workaround'' shows the percentage of vulnerabilities that have configuration workarounds. Column ``Period'' shows the earliest and latest time when the vulnerabilities are reported. For IE and Office, we cite the number of configuration options measured by Ocasta~\cite{Ocasta}. For other applications, we obtain the list of their configuration options from their source code using either static analysis, manual examination, or user documentation.  While it is difficult to say whether a small number of configuration options indicates that each option covers a large amount of code, in general we can see that the number of configuration options is usually small.  

We observe several trends in the results of our study.  First, configuration workarounds are listed for every application in our study.  This shows that the use and disclosure of configuration workarounds is widespread across software projects.  Second, the percentage of vulnerabilities that have workarounds is relatively low -- a weighted average (by \# of vulnerabilities) shows that only 25.2\% of the vulnerabilities have configuration workarounds.  As a result, the cases where a security vulnerability can be neutralized with an existing configuration workaround is quite uncommon.  

Qualitatively, we find that many configuration workarounds disable an entire ``module'' of functionality that was associated with the vulnerable code.  This suggests that many configuration workarounds cause some collateral damage; they not only disable the vulnerable code, but may also unnecessarily disable other non-vulnerable functionality.  For example, vulnerability CVE-2011-4362 in lighttpd~\cite{CVE-2011-4362-EXPLOIT} is the result of an incorrect bounds check in the code that is only called during base64 decoding of credentials for HTTP basic authentication.  However, the posted configuration workaround disables all types of authentication because it is the only configuration option that can prevent the vulnerable code from being executed.  This means that other types of authentication that do not rely on base64 decoding, such as digest and NTLM authentication, are needlessly disabled.  In general, the coarseness of the configuration options means that the configuration workarounds frequently disable more functionality than is strictly necessary.


Objectively speaking, it is not a complete surprise that configuration workarounds, while widespread in their usage across applications, are generally applicable to a minority of vulnerabilities and might only be able to disable code at a coarse granularity.  Having fewer configuration options simplifies testing and generally improves usability, motivating developers to minimize the configurability of their applications. There are likely many regions of code that cannot be disabled by the limited number of configuration options, resulting in many vulnerabilities for which there is no configuration workaround.

\section{Overview} \label{sec:overview}

\subsection{SWRR objectives}
From our study of configuration workarounds we found that while configuration workarounds are commonly used, they have very low coverage of vulnerabilities, thus reducing their utility.  Despite this, the reason why configuration workarounds are still used is that they impose no additional effort on the part of the developer. In essence, they provide a small, but tangible benefit for free. While it might seem obvious that a special purpose mechanism like SWRRs can improve on the coverage of configuration workarounds, we remain cognizant that to be competitive, they must at the same time impose little or no engineering cost. Furthermore, as a temporary alternative to a patch, they must be quick to generate as compared to construct a patch. We achieve low-effort by automatically generating SWRRs with a static analysis tool called Talos.  However, if designed improperly, an automatically generated SWRR may do more harm than a manually created configuration workaround.  As a result, we state the following objectives for our design of Talos and the SWRRs it creates:

\begin{itemize}
	\item {\bf Security:} An SWRR should neutralize its intended vulnerability and, in doing so, it should not introduce new bugs or vulnerabilities.  
	\item {\bf Effective Coverage:} SWRRs should be able to cover many more vulnerabilities than configuration workarounds.  Effective coverage is a product of two components: (1) the number of vulnerabilities whose code SWRRs can disable (which we call ``basic coverage''), and (2) the percentage of SWRRs that, when enabled, result in a minor loss of functionality similar to what would be expected from a configuration workaround.  
	\item{\bf Low Cost:} SWRRs are mechanically inserted into an application using Talos, thus minimizing the engineering effort required to use SWRRs.  In cases where a binary SWRR patch cannot be issued, it should be possible to perform ``in-place'' deployment of SWRRs, similar to deployment of configuration workarounds, and with minimal performance overhead.
\end{itemize}

Configuration workarounds are very unlikely to introduce new bugs or vulnerabilities since they have been tested; we expect the same behaviour from SWRRs, however, we limit our security objective to avoiding vulnerabilities that can compromise the confidentiality and integrity of a program.  It is possible and acceptable for Talos to create an SWRR that causes the application to terminate, even though this creates a potential denial-of-service vulnerability.  We believe this is acceptable because most state-of-the-art vulnerability mitigation techniques (such as Address Space Layout Randomization (ASLR), Control Flow Integrity (CFI), and non-executable stacks) aim to turn memory corruption exploits or malicious control flow transfers into program crashes, which also result in the termination of the program~\cite{shacham04, DEP, Abadi2005, Zhang2013CFI,Zhang2013CFICOTS,Tice2014}.  As a result, our design of SWRRs aims to completely avoid confidentiality and integrity vulnerabilities in exchange for some (small) probability of introducing a denial-of-service vulnerability.

A full patch requires at least the same or more effort to generate than an SWRR.  This is because a full patch must preserve all the functionality of the application while SWRRs explicitly allow some loss of functionality.  Specifically, to create a full patch, a developer needs to understand the exact cause of the vulnerability and all the conditions under which the vulnerability is triggered. In addition, the developer needs to design and implement new code that retains all desired functionality of the old code but does not contain the vulnerability.  In contrast, Talos only requires knowing the function in which the vulnerable code is located, which can usually be obtained from a crash report.  

The difference in effort is dependent on the complexity of the vulnerability.  The amount of effort to create a full patch generally increases as the complexity of the vulnerability increases. On the contrary, the amount of effort to generate an SWRR is essentially constant, as it just requires knowing the function that contains the vulnerable code, and the effort to get this information is independent of the complexity of the vulnerability. Consequently, the more complex the vulnerability is, the larger the difference in effort will be. For simple vulnerabilities, the difference in effort might be small, but the results in Table~\ref{tbl:patch_study} suggest that a fair number of vulnerabilities can be quite complex.

\subsection{SWRR deployment}

There are two possible deployment methods for SWRRs. In the first deployment method of SWRR, which we call {\em in-place deployment}, Talos is run on the application code base before it is released.  Talos inserts an {\em SWRR check} into every function in the application.  Each SWRR check reads and checks a corresponding {\em SWRR option} in an accompanying {\em SWRR configuration file}.  This allows the application developer to selectively disable code in an application without having to replace the binary by pushing out a new SWRR configuration file instead.  Alternatively, the user may change the configuration file to enable the appropriate SWRR if they know which function the vulnerability occurs in. In-place SWRR deployment is useful in scenarios where runtime performance is not critical or in scenarios where updating binaries is difficult, such as in smartphones or other embedded devices.  

In the second deployment method of SWRR, which we call {\em patch-based deployment}, the application developer will run Talos on the application code base when they learn of a new vulnerability, passing Talos the information it requires about the vulnerability.  Talos will then insert code that will disable the vulnerable function(s) and trigger error-handling code to return the application to a good state.  The application developer will then compile the instrumented code and issue the resulting binary as a temporary patch to users.  The application developer can perform minimal testing on the temporary patch as SWRRs are unlikely to cause serious loss of functionality in most cases, which is shown in our evaluation.

Using an SWRR requires that the location of the vulnerability be known. We argue that this is a reasonable requirement -- by the time a vulnerability is discovered and confirmed, the location of the vulnerable code is generally known, albeit a proof-of-concept exploit is often not publicly available.  For example, many of the CVE vulnerability reports we used in our experiments specifically list the function in which the vulnerability is located. 

Each of the two SWRR deployment methods has its own pros and cons. In-place deployment frees the developer from the need to re-compile the code and roll-out new code, but imposes a a slight increase of code size and minor performance overhead, as we will show in Section~\ref{sec:evaluation}. Patch-based deployment on the other hand has no code size or runtime overhead, but requires new binary code to be distributed and installed.

As the main goal of SWRR is to provide a rapid response when a vulnerability is newly discovered, we use Figure~\ref{fig:comparison} to illustrate the similarities and differences in the steps required for the two SWRR deployment methods and the conventional method of releasing a full patch. In the figure, the workflows of different approaches are distinguished with the use of different types of arrows. The legends used in the figure are explained in the dotted box at the bottom of the figure.

Regardless of which method is used to address the vulnerability, the discovery of a new vulnerability always starts with triage and finding the location of the vulnerability. After that, each method consists of different steps. First, releasing a patch requires software developers to find the cause of the vulnerability and to construct a patch, which can require a considerable amount of developer effort and time.  In addition, since full patches must not break existing functionality, regression tests must be performed.  Furthermore, the vendor must release the patch and end-users must install the patch.  Second, in-place SWRR deployment requires developers to identify the SWRR that can mitigate the vulnerability, which can be done by simply running Talos, and end-users to activate the SWRR by installing the new SWRR configuration file or enabling the appropriate SWRR configuration option.  Finally, patch-based SWRR deployment requires developers to generate an SWRR specifically for the vulnerability, which is also done by running Talos, vendors to release the SWRR as a patch, and end-users to install the SWRR patch. Note that at the end, the conventional approach of releasing a patch will fix the vulnerability, while both SWRR deployment methods only mitigate the vulnerability. However, both SWRR deployment methods require fewer steps and the steps that they require are simpler and less time-consuming than those of a full patch, due to the nature of their purposes and the aid of an automated tool like Talos.

\myfig{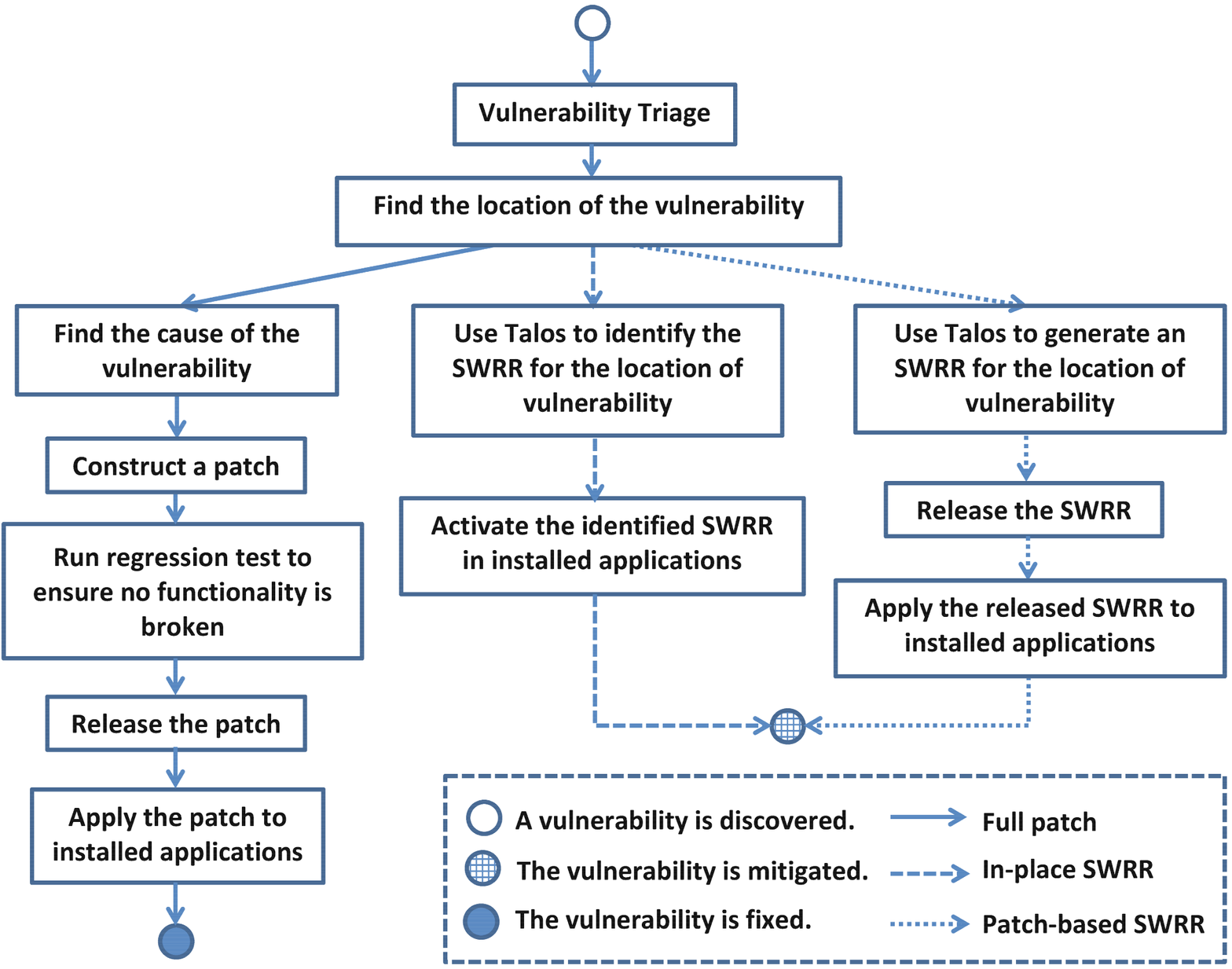}{The comparison of the different approaches to addressing a newly discovered vulnerability.}{fig:comparison}{1.0}{-10pt}

\subsection{The error-handling code intuition}

Talos must insert SWRRs that neutralize vulnerable functions without violating security,  Further, it must do this without needing to understand complex program-specific semantics.  As a result, Talos is almost completely application-agnostic, requiring only a small amount of application-specific information from developers. The key to enabling Talos to do this is to find an application characteristic that (1) is present and similar across nearly all applications, and (2) can allow Talos to recover from the unexpected redirection of execution to avoid vulnerable code.


Our intuition is that code whose purpose is to handle unexpected or abnormal error conditions fits these requirements.  First, error-handling is found in nearly every type of application.  Essentially any sufficiently complex application that interacts with its environment must gracefully handle unexpected situations such as invalid inputs, inadequate resources, or unexpected delays that it encounters; this is generally accomplished with what we generically refer to as {\em error-handling code}. Second, error-handling code is designed to be invoked when the application encounters these unexpected or abnormal situations and thus, by nature, it must conservatively return the application back to a known state.  In fact, the majority of error-handling code takes great pains to try to avoid violating confidentiality by leaking sensitive information or violating integrity by corrupting data.  Instead, most error-handling code remedies an abnormal situation by aborting the current task and cleaning up any intermediate state or, in the worst case, gracefully halting the application if continuation is not possible.  As a result, the intuition behind the goals of error-handling code fits well with the security goal of protecting the confidentiality and integrity of applications.

\lstdefinestyle{C}{
	xleftmargin=10pt,
	xrightmargin=0pt,
	basicstyle=\scriptsize\bfseries\ttfamily,
	showspaces=false,
	showtabs=false,
	breaklines=true,
	showstringspaces=false,
	breakatwhitespace=true,
	commentstyle=\color{mygreen},
	keywordstyle=\color{myblue},
	stringstyle=\color{myred},
	numbers=left,
	numbersep=5pt,
	numberstyle=\scriptsize\color{mygray},
	tabsize=2,
	frame=TB,
	framesep=5pt,
	escapeinside={\%*}{*},
	title=\lstname
}

\section{Talos} \label{sec:talos}
We now describe how Talos inserts SWRRs into application code without introducing new security vulnerabilities.  First, we explain how Talos sets about instrumenting an application with SWRRs.  Then, we detail the heuristics Talos uses to identify error-handling code within an application for the purposes of SWRR instrumentation. 

\subsection{Inserting SWRRs}\label{sec:insert_fpsw}

When designing Talos, we had to decide on the granularity of code that each inserted SWRR should enable or disable. The granularity of code that is protected by each SWRR has a bearing on its security and unobtrusiveness.  This is because error-handling code can broadly be classified into two categories: intra-procedural error-handlers that operate completely within a function, and inter-procedural error-handlers that are unable to completely handle the error within the function and must expose the error to the function's caller.  The error handlers in the former category are difficult for Talos to use as they are tightly coupled with the path within the function used to invoke the error-handling path.  For example, they may free memory that they know was allocated on the path leading to the error-handling code, or conversely fail to free memory since they know the paths leading to the error-handling code did not allocate it.  If Talos redirects execution to such an error-handling path without understanding the internal semantics of the function, it could result in a double-free bug.

However, inter-procedural error-handling code that exposes the error to the caller must be more conservative because it must be written in such a way that correctness guarantees are met independently of the calling context.  As a result, such error-handling code often seeks to ensure that modifications made to application state by the function are undone and that an appropriate value is returned to the caller so that the caller can then handle the failure.  For example, an input sanitization function that fails due to an out-of-memory error might free any resources acquired up to that point and then return an error code to the caller so that the caller can conservatively reject the unsanitized input.  This intuition implies that functions that contain such error-handling code can safely do nothing as long as the caller is notified that the function has encountered an error.  As a result, Talos instruments SWRRs to enable or disable code at the granularity of a function.  While there is no guarantee that this intuition is always true, we find that it does hold for a large number of cases allowing Talos to instrument applications with SWRRs that are secure and provide better effective coverage than configuration workarounds as we demonstrate in our evaluation in  Section~\ref{sec:evaluation}.  

Given that an SWRR option should control the execution of a function, instrumenting a function with an SWRR is fairly straightforward.  To instrument a function, Talos adds the code in Listing~\ref{lst:instrument} or Listing~\ref{lst:instrument2} to the function, depending on whether in-place deployment or patch-based deployment is used.  For in-place deployment, a check is first performed on line~\ref{lst:instrument:option} to determine whether the corresponding SWRR option ({\em SWRR\_option}) is enabled; if it is, the entire function body is skipped and the error code ({\em error\_code}) that has been statically extracted from the error-handling code is returned to the caller on line~\ref{lst:instrument:error_code_inplace}.  In this section, the text will mostly assume in-place deployment since it is the slightly more complex of the two options.

\lstset{emph={fpsw_option,error_code},emphstyle=\bfseries\it}
\begin{lstlisting}  [float,floatplacement=T,label=lst:instrument,language=C,style=C,caption=SWRR instrumentation - In-place Deployment\vspace{-18pt}]
int example_function(...) {
	/* SWRR inserted at top of function */
	if (SWRR_enabled(<SWRR_option>))  %*\label{lst:instrument:option}*
	  return <error_code>; %*\label{lst:instrument:error_code_inplace}*
	  
	/* original function body */
	...
}
\end{lstlisting}

\lstset{emph={fpsw_option,error_code},emphstyle=\bfseries\it}
\begin{lstlisting}  [float,floatplacement=T,label=lst:instrument2,language=C,style=C,caption=SWRR instrumentation - Patch-based Deployment\vspace{-18pt}]
int example_function(...) {
	/* SWRR inserted at top of function */
    return <error_code>; %*\label{lst:instrument:error_code_patch}*
	  
	/* original function body */
	...
}
\end{lstlisting}

\begin{figure}
	\begin{algorithmic}
		\Procedure{Find\_Functions}{$Functions$}
		\State $to\_instrument \gets \emptyset$
		\State $SWRR\_map \gets \emptyset$
		\For{$f \in Functions$}\Comment{Apply 2 main heuristics}
		\If{error\_logging(f)}
		\State $to\_instrument \gets \{f,error\_code(f)\}$
		\State $SWRR\_map \gets \{f,new\_option()\}$
		\State $remove(Functions, f)$
		\ElsIf{NULL\_return(f)}
		\State $to\_instrument \gets \{f,NULL\}$
		\State $SWRR\_map \gets \{f,new\_option()\}$
		\State $remove(Functions, f)$
		\EndIf
		\EndFor
		\For{$f \in Functions$} \Comment{Apply 2 extension heuristics}
		\If{$f' = $ propagate$(f,to\_instrument)$}
		\State $SWRR\_map \gets f,new\_option()\}$
		\State $to\_instrument \leftarrow \{f,error\_code(f')\}$
		\State $remove(Functions, f)$ 
		\EndIf
		\EndFor 
		\For{$f \in Functions$} 
		\If{$f' = $ indirect$(f,to\_instrument)$}
		\State $SWRR\_map \gets \{f, option(f')\}$
		\EndIf
		\EndFor
		
		\Return{$\{to\_instrument,SWRR\_map\}$}
		\EndProcedure
	\end{algorithmic}
	\caption{Talos algorithm for identifying functions to instrument.}
	\label{fig:talos_alg}
\end{figure}

Since a suitable error code must be found for each function instrumented with an SWRR, Talos can only instrument a function if: (1) it can determine if the function has inter-procedural error-handling code, and (2) it can extract the value that the error-handling code returns to be used as the error code.  While other work has used dynamic profiling to try to identify error-handling code~\cite{assure}, this requires a comprehensive suite of test inputs to find all error-handling code.  We assume this is not always available, so to maintain a low deployment cost, Talos relies exclusively on static analysis. Talos thus uses several heuristics based on common programming idioms that are indicative of error-handling code.  

The procedure Talos uses for deciding which functions in an application to instrument has several stages as illustrated in Figure~\ref{fig:talos_alg}.  The procedure takes as input the set of all functions in the application; it returns a set of functions capable of being instrumented as well as a map of functions to their corresponding SWRR options.  Talos first iterates over each function, applying the two main heuristics used to statically detect if the function has error-handling code.  If such code is detected, then Talos adds the function along with the {\em error\_code} extracted from the error-handling code to the set of functions it will instrument and removes it from further consideration.  In addition, Talos creates a new SWRR option for the function and adds it to the SWRR option-to-function map it maintains.  After all functions have been checked with the two main heuristics, Talos then applies the two ``extension'' heuristics to identify cases where it can extend error-handling code into a function's caller or callee.  Talos uses the {\em error propagation} heuristic to identify cases where the error code for a function can be used in an SWRR for the callers of the function, even if the callers themselves do not have error-handling code.  Finally, Talos also uses the {\em indirect} heuristic to identify any remaining cases where a function doesn't have error-handling code but can be disabled by a caller (or callers) that have been instrumented by an SWRR.  In these cases, the SWRR map is updated so that this function is also associated with the SWRR option of its caller(s).

\subsection{Main heuristics}\label{sec:talos:main}

We first describe the two main heuristics Talos uses to identify error-handling code in functions.  We will then describe the two extension heuristics.	


\subsubsection{Error-logging function heuristic} \label{sec:error}
The first heuristic is used to identify program paths that call error-logging functions.  Error-logging functions are called to log information when the application encounters an error.  To use this heuristic, Talos requires developers to specify the error-logging functions in an application. For each of the surveyed applications, Table~\ref{tbl:error} lists the total number of functions and the number of error-logging functions, where we have manually identified the latter by inspecting the source code.  We can see that, even in fairly large applications with hundreds or thousands of functions, many applications have very few and, in many cases, only one error-logging function.  Anecdotally, we also find that if there is more than one function, they are still often easy to find because they are all declared within a single header-file in the application source code.  Thus, we feel that the effort required for developers to specify the error-logging functions in an application is quite reasonable.  

\begin{table}
\begin{center}
		\caption{Number of functions and number of error-logging functions.}
	\begin{tabular}{|l|r|r|r|}
		\hline
		\textbf{App.} & \textbf {Functions} & \textbf{Error Funcs.}\\
		\hline
		lighttpd & 665 & 1 \\
		\hline
		apache &  2,082 & 4\\
		\hline
		squid & 1,346 & 1 \\
		\hline
		proftpd & 1,092 & 1\\
		\hline
		sqlite & 1,562 & 3 \\
		\hline
	\end{tabular}
	\label{tbl:error}
	\end{center}
\end{table}


The presence of an error-logging function is indicative of error-handling code.  However, recall that Talos requires the error-handling code to be inter-procedural, which means that it must also signal the error to the function's caller.  Thus, to identify such code using an error-logging function, Talos requires the following: (1) the error-handling code must call an error-logging function, (2) the error-handling code must return a constant value, and (3) the error-logging function and return statement must be guarded by a conditional branch.  A conditional check that dominates the error-logging function indicates that the path will only be taken under specific circumstances and a constant return value is required for the current function to signal to its caller that it terminated with an abnormal condition.  Listing~\ref{lst:typical error-handling code}, which shows error-handling code in Apache 2.2.19, illustrates how real code fits this heuristic. The error-handling code is only executed when condition {\tt (name == NULL)} is true. It then calls the logging function {\tt ap\_log\_error()} and returns the constant value {\tt APR\_EBADF} to its caller, which indicates that it cannot proceed because of a bad filename.  

Talos instruments such functions to always return a constant return value consistent with the error-handling code when the SWRR is activated (i.e. in place of {\em error\_code}) in Listing~\ref{lst:instrument}.

\begin{lstlisting}[float, language=C, style=C, caption=Error-logging code example from Apache\vspace{-18pt}, label={lst:typical error-handling code}]
if (name == NULL) {
	/* Apache's error logging function */
	ap_log_error(APLOG_MARK, APLOG_ERR, 0, NULL, "Internal error: pcfg_openfile() called with NULL filename");
	return APR_EBADF; /* indicates to caller that error occured */
}
\end{lstlisting}


\subsubsection{NULL return heuristic}
If the error-logging heuristic does not identify the presence of error-handling code, Talos next uses the NULL return heuristic.  The intuition behind this heuristic is that when a function that normally returns a pointer returns NULL instead, it indicates that the function could not successfully perform its normal operation.  This may happen due to an unexpected error or due to an invalid input.  

Talos would instrument such functions with an SWRR that returns NULL as its error code.  However, Talos must be conservative because not all functions can legally return a NULL to their callers.  If an SWRR were to force such a function to return NULL, the caller may dereference the value without checking for NULL and crash the program.  To infer whether a function can return NULL or not, Talos checks that there is at least one instance of a call to the function where the caller checks the return value against NULL.  The reason Talos does not do this for all call sites is that in some cases, the check for NULL may be hard to detect.  For example, consider the case where the caller writes the returned pointer value to a linked list and then the value is only checked against NULL when it is dequeued from the linked list.

\subsection{Extension heuristics}

We now discuss the two heuristics that Talos uses to extend coverage from functions that have identified error-handling code to those that do not.

\subsubsection{Error propagation heuristic}

This heuristic is based on the observation that many times the error code returned by a function is used as a return value by the caller of such functions.  This has the effect of propagating error codes up the call chain and, as a result, can be used to detect the correct error codes for both callees and callers of a function.


This error propagation manifests in three ways.  First, we find that some functions have an execution path that calls another function and simply uses the return value of the function call as their own return value. As illustrated by a simplified code snippet from lighttpd in Listing~\ref{lst:propagating_error_return}, \texttt{config\_insert\_values\_global} calls \texttt{config\_insert\_values\_internal} and uses the return value of the callee as its own return value at line~\ref{lst:propagating_error_return:direct}. As a consequence, the error code of \texttt{-1} for \texttt{config\_insert\_values\_internal}, identified by Talos using the error-logging heuristic at line~\ref{lst:propagating_error_return:error_code}, can be used as the error code for \texttt{config\_insert\_values\_global}. 

Second, the error code can also be translated before it is propagated up the call chain, as illustrated again in Listing~\ref{lst:propagating_error_return}.  Here, \texttt{mod\_secdownload\_set\_defaults} checks the return value of a call to \texttt{config\_insert\_values\_global} at line~\ref{lst:propagating_error_return:indirect} and returns a constant value {\tt HANDLER\_ERROR} at line~\ref{lst:propagating_error_return:indirect_return}  if the return value from {\tt config\_insert\_} {\tt values\_global} indicates an error. Unlike in the first case, \texttt{mod\_secdownload\_set\_defaults} does not use the return value of \texttt{config\_insert\_values\_global} directly, but translates it to its own error code if the callee returns an error. To identify this kind of error propagation, Talos looks for a statement that returns a constant and is control-dependent on the return value of a function call. Talos then checks: (1) whether the function in the predicate has been previously identified as having error-handling code, and (2) whether the identified error code can satisfy the predicate of the control dependency.  If so, the returned value becomes the error code for the function and Talos marks the function as eligible for SWRR instrumentation.  In the example, Talos identifies \texttt{HANDLER\_ERROR} as the error code for \texttt{mod\_secdownload\_set\_defaults}.

Third, the error code can be inferred down the call chain as shown in the code of \texttt{http\_request\_parse} in Listing ~\ref{lst:propagating_error_return}. \texttt{http\_request\_parse} has an error path that calls an error logging function when the return value of the call to \texttt{request\_check\_hostname} is not zero. From this, Talos infers that the error code of \texttt{request\_check\_hostname} must be a non-zero value. To identify this kind of error propagation, Talos checks if any identified error path is control dependent on the value of a predicate involving the return value of a function call. If it is, Talos tries to find a constant value that can satisfy the predicate and then uses that constant value as the error code of the callee of the function. For this example, Talos identifies {\tt1} as the error code for \texttt{request\_check\_hostname}.


\lstset{emph={ftpHtmlifyListEntry},emphstyle=\color{myred}\bfseries}
\begin{lstlisting}  [float,floatplacement=T,label=lst:propagating_error_return,language=C,style=C,caption=Error propagation  example from lighttpd\vspace{-18pt}]
int config_insert_values_global(....) {
....
   return config_insert_values_internal(....); %*\label{lst:propagating_error_return:direct}*
}

int config_insert_values_internal(....) {
....
   if (....) {
       log_error_write(....); // error logging
       return -1; %*\label{lst:propagating_error_return:error_code}*
   }
....
}

SETDEFAULTS_FUNC(mod_secdownload_set_defaults) {
....
   if (0 != config_insert_values_global(....)) { %*\label{lst:propagating_error_return:indirect}*
       return HANDLER_ERROR; %*\label{lst:propagating_error_return:indirect_return}*
   }
....
}

int http_request_parse(....) {
....
   if (0 != request_check_hostname(....)) {
       log_error_write(....); // error logging
       return 0;
   }
....
}
\end{lstlisting}

\subsubsection{Indirect heuristic}
If a function is only called by functions that have been identified as eligible for instrumentation, Talos takes advantage of the fact that by disabling all the callers of the function, the function itself can be disabled by SWRRs.  In these cases, Talos does not insert any instrumentation into these functions, but simply updates the SWRR map to indicate that the function in question can be disabled by activating one or more other SWRRs.

\section{Implementation}\label{sec:implementation}

\tikzstyle{decision} = [diamond, draw, fill=blue!20, 
    text width=4.5em, text badly centered, node distance=3cm, inner sep=0pt]
\tikzstyle{block} = [rectangle, draw, fill=blue!20, 
    text width=5em, text centered, rounded corners, minimum height=4em]
\tikzstyle{line} = [draw, -latex']
\tikzstyle{cloud} = [draw, ellipse,fill=red!20, node distance=3cm,
    minimum height=2em]

\myfig{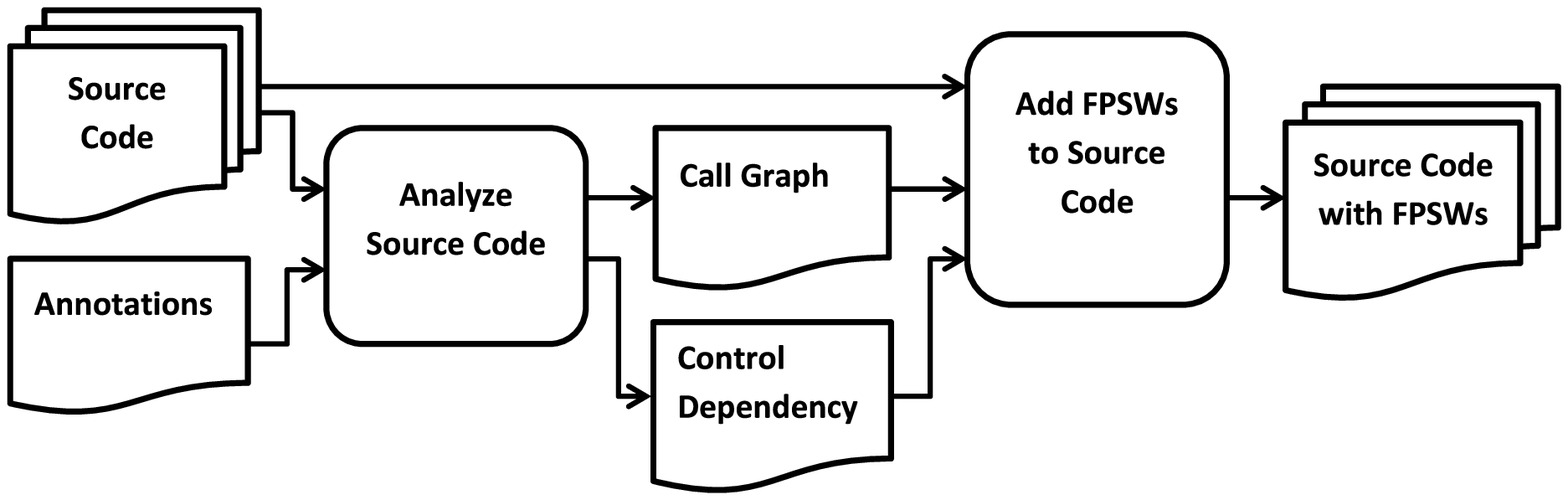}{Workflow of Talos}{fig:workflow}{1.0}{-10pt}


We have implemented a prototype of Talos. Due to the fact that Talos needs a program's call graph to find locations for SWRR insertion, our prototype instruments a program in two phases, as shown in Figure \ref{fig:workflow}. The first phase analyzes the source code of the program and is implemented as an analysis pass of LLVM using 1,823 lines of C/C++ code, while the second phase adds SWRRs to the source code and is implemented using 1,852 lines of Python code. In the first phase, Talos takes as input the source code of a program and the annotation of the error logging functions of the program, analyzes the source code using static analysis, and outputs: the program's call graph, the control dependency of each statement of the program, whether each statement is followed by a return, the start line number of each function, and the line number of each statement. In the second phase, it adds SWRRs to as many functions as possible in the source code based on the output of the first phase.

During our implementation, we found that function calls using function pointers are frequently used by applications, particularly to invoke the functionality of loadable modules. Loadable modules  are often used as a configuration workaround for vulnerabilities, so we expect that SWRRs should work for these as well.  We note that these kinds of function calls usually use function pointers embedded as fields of some C/C++ structures. To identify the caller and callee of a function call using a function pointer, we match a call to a function pointer field of a structure, by identifying the assignment or initialization of the same field. This method is imprecise, but we did not notice any issue with it in practice. 

To identify error-handling code that can be used for SWRRs, we need to find out whether a call to an error logging function is followed by a return statement. At first, we tried to label such cases when a call is followed by a return statement within the same basic block of the call. However, we found that LLVM merges all occurrences of return statements within a function into a single return at the end of the function and replaces all other return statements with branch statements. Sometimes a return is translated into a chain of unconditional branch statements that lead to the only return statement of a function. Hence a call to an error logging function and the return statement following it sometimes do not belong to the same basic block. Furthermore, some applications' error logging function is actually a macro defined as an {\tt if} statement, so the call to the error logging function and the return statement belong to two different basic blocks. As a consequence, we label a call as being followed by a return when the call is on a path that unconditionally leads to a return statement.



\section{Evaluation}\label{sec:evaluation}

We evaluate how well SWRRs created by Talos meet the three objectives we laid out in Section~\ref{sec:overview}.   First, we evaluate the security of SWRRs by testing them against 11 real-world vulnerabilities in five popular applications: two web servers, a web cache/proxy, an ftp server, and one database application.  We then evaluate the effective coverage by measuring the basic coverage of SWRRs and the rate of unobtrusive SWRRs.  We define ``unobtrusive SWRR'' as an SWRR that only disables minor application functionality while leaving the majority of an application's functionality intact, much like a configuration workaround, and an ``obtrusive SWRR'' as an SWRR that disables the majority of an application's functionality, making it unusable. Thus, the basic coverage of SWRRs is reduced to their effective coverage by the percentage of SWRRs that are obtrusive. Finally, we evaluate the performance cost of SWRRs when using in-place deployment. 

All our evaluations were conducted on a 4-core 3.4GHz Intel Core i7-2600 workstation, with 16GB RAM, 3TB of SATA hard drive and running 64-bit Ubuntu 12.04.


\subsection{Security}\label{sec:security}
This evaluation answers the question: \textbf{``Do SWRRs successfully neutralize vulnerabilities without introducing new vulnerabilities?"} To test an SWRR, we need one vulnerability that is covered by the SWRR; to check whether an SWRR neutralizes a vulnerability, we also need an exploit for that vulnerability. These two requirements limit the number of SWRRs that can be evaluated in detail, and it also requires non-trivial manual effort to check whether new vulnerabilities are introduced. Nevertheless, we make our best effort to find as many vulnerabilities as possible that could be used for this evaluation. The resulting 11 vulnerabilities, disclosed between 2010 and 2015, are used to evaluate the five popular applications as shown in Table~\ref{tbl:vulnerabilities}.   

To validate security, we check if the SWRR neutralizing the vulnerability successfully thwarts an exploit of the vulnerability. To test whether the exploit is neutralized or not, we either use a published exploit or create a proof-of-concept exploit if no published exploit is available. We verify that the exploit works on the unprotected application and then enable protection using the appropriate SWRR option and try the exploit again.  If the exploit fails, we say the SWRR has protected the security of the application.  

We can also test whether an SWRR is unobtrusive or not.  To do this, we first classify the functionality of each application into two categories, major and minor, by studying its user documentation. We then design two sets of test inputs, major and minor, to exercise as much the application's major functionality and minor functionality as possible. For each application, we make use of the existing test suite of an application if such a test suite is available. Otherwise we make our best effort to create our own sets of test inputs and test suite. We then use this test suite to determine if no or only minor functionality is lost, in which case the SWRR is unobtrusive; if major functionality is also lost, the SWRR is obtrusive.


Our results are summarized in Table~\ref{tbl:vulnerabilities}, which also gives the heuristic used to instrument the SWRR that neutralizes the vulnerability, as well as whether availability is violated. Column ``Security?'' shows whether the exploit against a vulnerability is successfully neutralized by SWRR without introducing new vulnerabilities. Column ``Unobtrusive?'' shows whether the SWRR is unobtrusive. SWRRs successfully neutralize the exploits for all 11 vulnerabilities and in 8 cases there is no or only minor loss of functionality, making these SWRRs unobtrusive.  We provide details on all 11 cases below.  For the 3 cases where a posted configuration workaround also exists for the vulnerability, we compare the SWRRs unobtrusiveness with that of the configuration workaround.

%


\begin{table}
	\caption{Security of SWRRs.}
	\begin{center}
		\begin{tabular}{|l|l|p{1.2cm}|p{1cm}|l|}
			\hline
			\textbf{App.} & \textbf{CVE ID} & \textbf{Heuristics} & \textbf{Security?} & \textbf{Unobtrusive?}\\
			\hline
			lighttpd & CVE-2011-4362 & NULL Return & Yes & Yes\\
			\hline
			lighttpd & CVE-2012-5533 & Indirect & Yes & No\\
			\hline
			lighttpd & CVE-2014-2323 & Error-Propagation & Yes & No\\
			\hline
			apache & CVE-2014-0226 & Error-Logging & Yes & Yes\\
			\hline
			squid & CVE-2009-0478 & Indirect & Yes & No\\
			\hline
			squid & CVE-2014-3609 & Error-Logging & Yes & Yes\\
			\hline
			sqlite & CVE-2015-3414 & Error-Propagation & Yes & Yes\\
			\hline
			sqlite & OSVDB-119730 & Error-Logging & Yes & Yes\\
			\hline
			proftpd & OSVDB-69562 & Error-Propagation & Yes & Yes\\
			\hline
			proftpd & CVE-2010-3867 & Error-Logging & Yes & Yes\\
			\hline
			proftpd & CVE-2015-3306 & Error-Logging & Yes & Yes\\
			\hline
		\end{tabular}
		
		\label{tbl:vulnerabilities}
	\end{center}
\end{table}

\scbf{lighttpd - CVE-2011-4362} This vulnerability allows a remote attacker to cause an out-of-bounds memory error~\cite{CVE-2011-4362-EXPLOIT}. The function {\tt base64\_decode} takes an untrusted {\tt char*} and performs a base 64 decode during HTTP basic authentication by using each character in the untrusted string as a lookup into a table in memory.  As {\tt char*} is signed, an attacker could specify negative values and read memory from outside of the table.  {\tt base64\_decode} has error-handling code that returns NULL, so Talos instruments the function with an SWRR that returns NULL, which successfully neutralizes the vulnerability.  Since base64 decoding is disabled, all requests for basic HTTP authentication fail as if the password failed to decode properly.  However, lighttpd functions completely normally (including other forms of authentication) as long as basic HTTP authentication is not used.  This imposes less loss of functionality than the posted configuration workaround, which disables all forms of authentication.  Thus, Talos provides security and provides an unobtrusive SWRR for the vulnerability.

\scbf{lighttpd - CVE-2012-5533} This vulnerability allows a remote attacker to cause an infinite loop via a specially crafted HTTP connection header. The function {\tt http\_request\_split\_value} splits the fields in an HTTP connection header into an array, but can get into an infinite loop due to the vulnerability.  {\tt http\_request\_split\_value} does not have error-handling code, but its caller does have error-handling code that returns {\tt0}; Talos instruments the caller and successfully neutralizes the vulnerability, however, the side-effects of this are severe, as it causes all HTTP requests to be denied, because the caller is the main function that processes HTTP requests. As a result, while the SWRR provides security, because the SWRR is enabled for all HTTP requests, lighttpd is unable to respond to any HTTP request so there is a major loss of functionality.

\scbf{lighttpd - CVE-2014-2323} This vulnerability allows a remote attacker to execute an arbitrary SQL command via a specially crafted hostname in the host header of an HTTP request. The vulnerable function {\tt request\_check\_hostname} checks the validity of hostnames, but it fails to deny hostnames that contain SQL commands.  The caller of the function has an error path that calls an error logging function when the return value of the function is not zero, so Talos instruments the function with an SWRR that returns {\tt1}, which successfully neutralizes the vulnerability.  As a side-effect of activating the associated SWRR, any HTTP request that specifies a hostname (as opposed to an IP address) will receive a  ``400 - Bad Request'' error response. While the SWRR provides security, because the vulnerable code is used for all HTTP requests with a hostname, which is in most cases the vast majority of requests, there is a major loss of functionality.

\scbf{apache httpd - CVE-2014-0226} A race condition in the {\tt mod\_status} module allows an attacker to retrieve sensitive information~\cite{CVE-2014-0226-EXPLOIT}. The function {\tt status\_handler} displays administrative information about a web server, such as the web server's performance and overhead, as a web site. It does not synchronize the use of data that can be modified concurrently by a different thread. {\tt status\_handler} has error-handling code that calls an error logging function and returns {\tt HTTP\_INTERNAL\_SERVER\_ERROR},  so Talos instruments the function with an SWRR that returns the error code, which successfully neutralizes the vulnerability.  As a side-effect, all requests to the {\tt mod\_status} module return an error because {\tt status\_handler} is called in response to all requests to the module, but the application will continue to execute and respond to other requests normally.  This vulnerability has a posted configuration workaround, which disables the entire {\tt mod\_status} module, with the exact same loss of functionality as Talos' automatically generated SWRR.  As a result, Talos provides security with an unobtrusive SWRR for this vulnerability.

\scbf{squid - CVE-2009-0478} An integer overflow vulnerability allows a remote attacker to cause a denial-of-service by sending an HTTP request with a crafted HTTP protocol version number~\cite{CVE-2009-0478}. The function {\tt httpMsgParseRequestLine} converts the HTTP version number of an HTTP request from a string to an integer, but it uses a signed integer to store the converted version number.  As a result, a very large version number will cause an integer overflow and crash the server. {\tt httpMsgParseRequestLine} does not have error-handling code, but its caller does (returns NULL); Talos instruments the caller, which successfully neutralizes the vulnerability.  However, the side-effects of this are severe, as it causes all HTTP requests to be denied as every request must be parsed by {\tt httpMsgParseRequestLine} and calls to this function always generate an error with the SWRR enabled.  While the SWRR provides security, because the vulnerable code is used for all HTTP requests, squid is unable to respond to any HTTP request so there is a major loss of functionality.


\scbf{squid - CVE-2014-3609} A missing validity check on the byte range specification of an HTTP request allows a remote attacker to cause a denial-of-service by sending an HTTP request with a specially crafted byte range specification~\cite{CVE-2014-3609-WORKAROUND}. The function {\tt httpHdrRangeSpecParseCreate} parses the byte range specification of HTTP requests, but it does not correctly check the validity of the length calculated from certain byte range specifications and can cause the server to crash. {\tt httpHdrRangeSpecParseCreate} has error-handling code that calls an error logging function and returns NULL, so Talos instruments this function with an SWRR that returns NULL, which successfully neutralizes the vulnerability.  This causes the server to ignore the byte range specification from the client and always serve the full-length of the content.  No confidential information is leaked since the client would have received the full-length content anyways if it had not specified a byte range.  This vulnerability has a posted configuration workaround, which implements a filter that rejects requests with suspicious byte ranges.  The loss of functionality is similar to the SWRR -- only requests that specify byte ranges are affected in either case.  Talos preserves security in this case with an unobtrusive SWRR.  

\scbf{sqlite - CVE-2015-3414} A vulnerability in the code that parses collation-sequence names in SQL commands allows an attacker to cause memory corruption. The function {\tt sqlite3ExprAddCollateString} allocates memory for parsed collation-sequence names, but may use uninitialized memory when parsing a specially crafted collation-sequence name. {\tt sqlite3ExprAddCollateString} does not have error-handling code and simply uses the return value of function {\tt sqlite3ExprAddCollateToken} as its own return value. Due to imprecise static analysis, Talos incorrectly identifies that {\tt sqlite3ExprAddCollateToken} could return NULL, although it is carefully written to always return a valid pointer. As a consequence, Talos instruments the function with an SWRR that returns NULL. Since {\tt sqlite3ExprAddCollateString} should not be able to return NULL, the caller does not check the return value before dereferencing it causing sqlite to crash.  If collation is not used, sqlite continues to operation normally, and since collation is not part of the core functionality of sqlite, we call this a minor loss of functionality.  If restarted, sqlite continues to function normally.



\scbf{sqlite - OSVDB-119730} An attacker can cause a memory error in sqlite with the meta command {\tt trace}, which turns on or off the tracing of the execution of commands. The function {\tt do\_meta\_command} processes all meta commands, which allows users to specify different settings when executing commands. It does not set a pointer to NULL after the memory which it references has been deallocated, and thus can cause a use-after-free memory error. {\tt do\_meta\_command} has error-handling code that calls an error logging function and returns {\tt1}, so Talos instruments the function with an SWRR that returns {\tt1}; this causes sqlite to return an error to the meta command request.  As a result, Talos protects the security of sqlite against this vulnerability.  However, because {\tt do\_meta\_command} is disabled, all other meta commands will also return an error, and thus the availability of all meta commands is violated.  However, because this is only confined to meta commands, which are not part of the core functionality of sqlite, this SWRR is unobtrusive.

\scbf{proftpd - OSVDB-69562} A backdoor that allows a remote attacker to access a root shell was planted into the source code of ProFTPD when ProFTPD's FTP server and mirrors were compromised~\cite{OSVDB-69562}. The backdoor was added to function {\tt pr\_help\_add\_response}, which creates responses to {\tt HELP} command, so that a {\tt HELP} command with a specific argument would cause ProFTPD to execute a shell that can be accessed remotely. The caller of the function has error-handling code that calls an error logging function when the return value of the function is not zero, so Talos instruments the function with an SWRR that returns {\tt1}, which successfully neutralizes the vulnerability.  As a result, the security and availability of the application are preserved.  However, as in the previous case, ProFTPD will respond to all {\tt HELP} commands with the error message ``Unknown command" thus impacting the availability of the {\tt HELP} facility.  However, all other FTP commands continue to function normally.  As a result, this is considered an unobtrusive SWRR.

\scbf{proftpd - CVE-2010-3867} Multiple vulnerabilities in the \texttt{mod\_site\_misc} module allow a remote attacker to perform various directory and file operations using \texttt{mod\_site\_misc} commands without authentication. All vulnerable functions, such as {\tt site\_misc\_mkdir} that creates a directory on the server upon users' requests, have error-handling code that calls an error logging function and returns NULL; Talos instruments each of these functions with an SWRR that returns NULL and when all of the SWRRs corresponding to these functions are enabled, ProFTPD returns an error for all the vulnerable \texttt{mod\_site\_misc} commands. Other than this side-effect, users can continue to use all other FTP commands and thus the SWRRs provide security and are unobtrusive.


\scbf{proftpd - CVE-2015-3306} Multiple vulnerabilities in the {\tt mod\_copy} module allow a remote attacker to read and write arbitrary files with {\tt mod\_copy} commands  without authentication. Similar to CVE-2010-3867, all vulnerable functions (such as {\tt copy\_copy}, which copies files between different locations on the server) have error-handling code that calls an error logging function and returns NULL; Talos instruments each of these with an SWRR that returns NULL when enabled.  Again, when the SWRR is activated, ProFTPD returns an error in response to all the vulnerable {\tt mod\_copy} commands. There are no other side effects and ProFTPD continues to work as expected, thus the SWRR provides security and is unobtrusive.

\begin{table}
	\caption{Basic coverage of SWRRs.  }
	\label{tbl:coverage}
\begin{center}
\begin{tabular}{|l|l|l|l|l|l|}
\hline
\textbf{App.} & \textbf{Protected} & \textbf{Logging.} & \textbf{Pointer.} & \textbf{Prop.} & \textbf{Indirect} \\
\hline
lighttpd & 89.8\% & 23.6\% & 1.5\% & 17.6\% & 47.1\% \\
\hline
apache & 77.5\% & 14.0\% & 11.9\% & 20.7\% & 30.9\% \\
\hline
squid & 76.6\% & 18.1\% & 5.6\% & 6.3\% & 46.4\% \\
\hline
proftpd & 86.1\% & 32.7\% & 13.6\% & 12.9\% & 26.9\% \\
\hline
sqlite & 45.3\% & 2.0\% & 6.5\% & 14.4\% & 22.4\% \\
\hline
\textbf{AVERAGE} & 75.1\% & 18.1\% & 7.8\% & 14.4\% & 34.7\% \\
\hline
\end{tabular}

\end{center}
\end{table}

\subsection{Effective coverage}

In this section, we aim to answer the question {\bf ``What is the percentage of vulnerabilities that can be mitigated with an unobtrusive SWRR?''}  To answer this question, we perform a quantitative measurement of the two components that make up the effective coverage of SWRRs: the basic coverage and the rate of unobtrusive SWRRs.


\scbf{Basic Coverage} To evaluate basic coverage, we measure the number of functions where Talos can find an error-handling path and identify an error-handling code to return, which is used to insert an SWRR.  This measurement across the five applications is shown in Table~\ref{tbl:coverage}.  The first ``Protected'' column shows the total percentage of functions that are protected by SWRRs in each application.  The remaining four columns then provide a breakdown by the percentage of functions that are protected by each of the four heuristics.  If we assume that potential vulnerabilities are uniformly distributed across functions in the application, then the percentage in the Protected column gives the basic coverage for the application, which is the likelihood that a potential vulnerability can be disabled by an SWRR.  

As Talos uses error-handling to infer the value that should be returned by an activated SWRR, the coverage depends very heavily on how much error-handling code is present in the application and how well Talos' heuristics can identify the error-handling code.  Among the five applications, sqlite has the lowest basic coverage of 45.3\% as well as a very low percentage of error-logging paths. In addition, sqlite has the lowest percentage of functions that can be protected indirectly. This is likely because sqlite has a simpler call graph than the other applications. 

On the other hand, lighttpd has the highest basic coverage of 89.8\% because it has a particularly high percentage of error logging paths as well as a high percentage of functions that can be protected indirectly. Unlike lighttpd, proftpd (the application that has the second highest coverage) has a high percentage of error-logging paths and NULL-returning functions, but has a lower percentage of functions that can be protected indirectly.

Overall, we can see that Talos has a basic coverage of 75.1\% across all applications and that each technique used by Talos plays an essential role in achieving the high coverage, although each one might have a different impact on the coverage for different applications. We also find that the majority of the functions can be directly protected by Talos.


\label{sec:unobtrusiveness}
\scbf{Rate of unobtrusive SWRRs} We wish to evaluate the unobtrusiveness of SWRRs over a large number of SWRRs.  To do this, we perform an experiment where we enable a large number of SWRRs and test whether they result in minor, major, or no loss of functionality.  To make it easy to test a large number of SWRRs, we instrument each application for in-place deployment so that we can activate each SWRR simply by changing configurations.  To ensure that all the SWRRs under our test are indeed executed, we first find out which functions are executed for the major and minor functionality test inputs used in Section~\ref{sec:security}, and then randomly choose approximately 25\% of the SWRRs corresponding to the executed functions to focus on in the interests of time.  In total we choose 320 SWRRs across all of the applications, as shown in Table~\ref{tbl:sec:unobtrusiveness}.  We then individually enable each of the selected SWRRs and run the test suite for the application. If the application passes both sets of test inputs or passes the major test inputs but fails the minor test inputs, we consider that the SWRR is unobtrusive. Otherwise, we consider the SWRR is obtrusive. 

\begin{table}
	\caption{Rate of unobtrusive SWRRs.}
	\begin{center}
		\begin{tabular}{|l|r|r|}
			\hline
			\textbf{App.} & \textbf{\#SWRRs} & \textbf{Unobtrusive} \\
			\hline
			lighttpd & 40 & 70.0\% \\
			\hline
			apache & 85 & 88.2\% \\
			\hline
			squid & 65 & 69.2\% \\
			\hline
			proftpd & 90 & 64.4\% \\
			\hline
			sqlite & 40 & 55.0\% \\
			\hline
			\textbf{AVERAGE} & 64 & 71.3\% \\
			\hline
		\end{tabular}
		\label{tbl:sec:unobtrusiveness}
	\end{center}
\end{table}

The results are tabulated in Table~\ref{tbl:sec:unobtrusiveness}. Column ``\#SWRRs'' shows the number of tested SWRRs for each application. Column ``Unobtrusiveness'' shows the percentage of tested SWRRs that are unobtrusive. A weighted average shows that 71.3\% of the SWRRs tested are unobtrusive, and thus preserve the major functionality of the application.  No application had a rate of unobtrusive SWRRs below 50\% indicating that the majority of SWRRs are unobtrusive.  

While one might believe that the rate of unobtrusive SWRRs is a function of the choice to use SWRRs to disable entire functions or the use of indirect protection, our analysis of some of the results indicates that this is not a major factor.  Rather, if the vulnerability is located in the core functionality of an application, it is unlikely that disabling code, even at a finer granularity, will preserve the major functionality of the application. Thus, the main factor for unobtrusiveness is the location of the vulnerability, which is out of Talos' control.  Essentially, our findings indicate that commonly executed code tends to have a higher rate of error-handling code, meaning there are more SWRRs located in commonly executed code with major functionality.  


In combining the average basic coverage with the average rate of unobtrusive SWRRs, we arrive at an effective coverage of 53.5\%, which gives the percentage of potential vulnerabilities that have an unobtrusive SWRR.  This is a significant 2.1$\times$ improvement over the 25.2\% coverage currently offered by configuration workarounds.


\subsection{Overhead}
When SWRRs are instrumented for in-place deployment, they can incur runtime overhead because they will check whether their corresponding configuration is activated at runtime every time the function into which they are instrumented is executed. When SWRRs are instrumented for patch-based deployment, there is no additional runtime overhead because there is no such check. Table~\ref{tbl:overhead} gives the overhead of SWRRs for in-place deployment, measured by the number of lines of source code added by Talos and the number of corresponding source files modified by Talos. Column ``App.'' shows the name of the application. Column ``LOC'' and ``Files'' show the number of lines of code and the number of original source files, respectively. Column ``Added LOC'' shows the percentage of the lines of source code added by Talos, and column ``Modified Files'' shows the percentage of corresponding source files modified by SWRRs. Column ``Runtime'' shows the runtime performance overhead of SWRRs. The last row shows the average for all columns. 

\begin{table}
	\caption{Overhead of SWRR.  }
	\label{tbl:overhead}
	\begin{center}
		\begin{tabular}{|l|R{1cm}|R{1cm}|r|R{1cm}|r|}
			\hline
			\textbf{App.} & \textbf{LOC} & \textbf{Added LOC} & \textbf{Files} & \textbf{Modified Files} & \textbf{Runtime}\\
			\hline
			lighttpd & 46,792 & 1.9\% & 79 & 92.4\% & 0.6\% \\
			\hline
			apache & 135,856 & 2.2\%  & 191 & 75.9\% & 2.3\% \\
			\hline
			squid & 70,407 & 2.4\% & 119 & 84.0\% & 1.5\% \\
			\hline
			proftpd & 69,808 & 2.9\% & 64 & 93.8\% & 1.2\% \\
			\hline
			sqlite & 153,020 & 0.8\% & 2 & 100\% & 1.0\% \\
			\hline
			\textbf{AVERAGE} & 95,176 & 2.0\% & 91 & 89.2\% & 1.3\% \\
			\hline
		\end{tabular}
		
	\end{center}
\end{table}

On one hand, we can see that Talos adds on average 2\% more lines of source code to implement SWRRs in applications. Given the high coverage achieved by Talos, this indicates that Talos has a very small footprint for each SWRR. On the other hand, the percentage of source files changed by Talos in order to add SWRRs is on average 89\%. This indicates that the functions protected by SWRRs are distributed among most of the source files. 

To measure the runtime performance overhead of SWRRs, we use standard benchmarks for each application if a standard benchmark is available, otherwise we write our own benchmark. For each application, we compare the performance of a version of the application that is hardened by SWRRs with a version that is not. We run each benchmark three times for each application and use the average of the three measurements. To have a fair comparison, we run the hardened version of each application with all SWRRs disabled, which has the same functionality of the original application but with the added execution of the SWRRs. 

For web servers including lighttpd and apache, we use ApacheBench~\cite{HTTPBench}. For the squid cache proxy, we also use ApacheBench, but we enable the use squid as web proxy in its settings. We use the throughput as the performance metric for these three applications. Roughly SWRRs reduce their throughput by 2\%.

For ftp servers including proftpd, we use the ftp benchmark included with pyftpdlib~\cite{pyftpdlib}, which measures the transfer rate for both file uploads and downloads. SWRRs reduce the transfer rate for file uploads by only 1.2\%, and have a negligible impact on file downloads.

For sqlite, we created our own benchmark,  which is based on the description of a series of SQL database performance tests on sqlite's official web site ~\cite{SQLiteBench}. It consists of over 70,000 SQL commands to create table, drop table, insert data, update data, query data, delete data, and perform database transactions. The benchmark measures the total execution time of all these SQL commands on sqlite database tables containing from 10,000 to 25,000 records of data. SWRRs incur a performance overhead of 1.0\% on sqlite.  On average in-place deployment of SWRRs has a very small runtime performance overhead of 1.3\% for all five applications.

\section{Discussion} \label{sec:discussion}



We begin by discussing the the limitations of SWRRs and then other operational issues associated with the deployment of SWRRs.

\subsection{Limitations}

The ability of SWRRs to neutralize vulnerabilities without security violations is limited by the assumption that applications correctly implement error-handling code.  Naturally, this is not the case -- applications developers may fail to identify and handle errors, or even if they do handle them, they may handle them incorrectly, as previous work has shown~\cite{remzi:eio,yuan:aspirator}.  Unfortunately, there is little that Talos can do if the error-handling code it calls contains bugs.  We hope, as previous work has also implored, that developers should pay more attention to the correctness of error-handling code.  While it is not invoked very often, when unexpected errors arise error-handling code is the last line of defense the application has against catastrophic failures.  

Another obvious limitation is that Talos has no control over where vulnerabilities occur.  As illustrated in lighttpd (CVE-2012-5533) and squid (CVE-2009-0478), if the vulnerability occurs in a key function that is used in many operations, then the availability of the application will be severely impacted.  Fortunately, this appears to be the less common case (only 3 out of 11 cases in our experiments).  We speculate that this is likely due to bugs and vulnerabilities occurring in less commonly executed code, as that code receives fewer opportunities for testing and has less chance of having a bug triggered in production use.

Currently Talos does not leverage the structured exception handling that is used in programming languages such as C++ and Java. However, Talos can be easily extended to do so, since exception handling makes error-handling code explicit, making it even easier for Talos to locate and use error-handling code in the application. In these cases, Talos can look for a type of exception that can be safely used to abort the execution of a function and generate an SWRR that throws the exception as the mechanism to prevent the execution of the function. To identify the exception that can be used, Talos can examine which exception is caught by existing exception handlers in the function or which exception is thrown by the function. If Talos cannot locate this kind of exception in the function itself, it can look for it in the callers of the function. We believe that leveraging structured exception handling would be interesting future work to explore.

\subsection{Other Issues}

Another question is whether SWRRs and their use can decrease the security of an application in other ways, or whether the SWRRs themselves can be circumvented by an attacker even when activated.  For example, even if the user activates an SWRR, an attacker can still corrupt the value of an SWRR option and re-enable the vulnerable code.  While this is possible, we believe it sufficiently raises the bar for the attacker, as she must have a memory corruption vulnerability that is not in the function(s) disabled by the activated SWRR(s).  In other words, to exploit an SWRR, the attacker needs a zero-day memory corruption vulnerability.  Given the nature of most memory corruption vulnerabilities, it would be likely that an attacker who has access to such a vulnerability would just use it to compromise the application directly rather than use it to disable an SWRR.  

In the rare instance that a memory corruption vulnerability doesn't allow remote code execution but can still corrupt an SWRR option, the attacker now has the ability to activate or deactivate SWRRs, allowing them to re-enable disabled functions or disable enabled ones.  As discussed above, they could thus silently re-enable vulnerabilities, or they could prevent code from being executed if the application has no known vulnerabilities.  However, as we have shown in this paper, activated SWRRs generally do not cause security vulnerabilities, and only impact availability.  Thus, the most the attacker can do is to cause a denial of service attack with a memory corruption vulnerability -- which is something they could likely already do with a memory corruption vulnerability even in the absence of SWRRs.


\section{Related Work}\label{sec:related}

Most closely related to the concept of an SWRR are proposals that attempt to mitigate security vulnerabilities or software flaws by altering the execution of an application.  These can be broken down into those that harden the application code and those that filter inputs to the application.

\scbf{Hardening application code} Systems that harden application code to prevent an attacker from exploiting vulnerabilities are a rich area of research.  For example, Software Fault Isolation (SFI)~\cite{sfi} and similar techniques~\cite{native,rocksalt}, instrument memory operations with bounds checks to make sure even erroneous ones cannot corrupt memory.  Another approach is to validate every control transfer with Control Flow Integrity (CFI)~\cite{Abadi2005, Niu2013, Niu2014, Zhang2013, Zhang2013CFICOTS, Criswell2014, Tice2014}.  Compared to the code instrumentation Talos uses for SWRRs, the code instrumentation that these systems use is either more complex in the case of CFI or needs to be called more frequently in the case of SFI.  As a result, these hardening approaches generally have a higher performance overhead.  

\scbf{Filtering inputs}
An alternative to hardening application code is to detect and filter malicious inputs.  In general, these techniques perform analysis of the application source code to generate a vulnerability-specific input filter that will detect inputs that could reach the specified vulnerability.  Some proposals detect and drop such inputs~\cite{Costa2007, Long2014-InputFilter, Susskraut2007, Wang2004}, while others convert malicious inputs into benign inputs~\cite{Long2012, Rinard2007}.  

For example, Bouncer uses static analysis, combined with dynamic symbolic analysis, on programs to infer the conditions that inputs must satisfy to exploit a vulnerability and then craft filters based on these conditions~\cite{Costa2007}.   HEALERS protects library functions by generating wrappers on them, which intercept malicious inputs and return an error condition instead of executing the vulnerable function~\cite{Susskraut2007}. It uses static analysis-guided fault injection to infer predicates on the input arguments to a function that can cause the function to crash.  HEALERS only works on libraries with a well defined error specification in their API.  In contrast, Talos works on arbitrary internal functions in an application, and thus must infer error paths and values since it does not assume they are specified.  Shields~\cite{Wang2004} uses statically extracted information to generate network filters, which then drop the network packets that might potentially trigger a vulnerability.  
Finally, SOAP heuristically converts malicious inputs into benign inputs \cite{Long2012} so that an application can still return partial (though sometimes inconsistent) results. It uses offline-training on benign inputs, along with input format supplied by its users, to infer critical fields of the input and the constraints over these critical fields. Based on these constraints, it rectifies inputs whose critical fields contain values that violate the constraints. 


The major difference between Talos and these approaches is that they all require some malicious input, i.e. proof-of-concept exploit, that can trigger a vulnerability. However, we find that most of times a proof-of-concept exploit is not publicly available for a disclosed vulnerability, probably due to security concerns. On the contrary, Talos does not require a proof-of-concept exploit and requires only the name of a vulnerable function, which is usually publicly available in vulnerability databases. 

\scbf{Resuming execution after faults}
Another area of related work tries to improve fault tolerance by allowing an application to continue execution after a fault has occurred~\cite{Gao2009, Rinard2004, Long2014-RecoveryShepherding, sidiroglou04using, Perkins2009, Carzaniga2013}.  In general, these do not have the same level of security as Talos as they cannot guarantee that the recovered application is secure, but they follow the same principle of detecting an erroneous application state and redirecting it to some non-erroneous state that Talos uses with SWRRs.

Failure-Oblivious Computing is proposed to improve the resilience of server applications after an attack has triggered memory errors, by augmenting an application to ignore memory errors~\cite{Rinard2004}. For out-of-bounds memory writes, it simply discards them. For out-of-bounds memory reads, it redirects them to a preallocated buffer that contains pre-defined values that are likely to reduce the possibility of a crash or infinite loop.  Recently, this work was followed by RCV, which further limits the propagation of the manufactured values within an application by skipping any system call that tries to use them~\cite{Long2014-RecoveryShepherding}.  Like Talos, these approaches seek to trigger error-handling code in the application.  The main difference is that these approaches are simpler in that they guess the values that will cause this to happen, while Talos uses static analysis on the application source code to discover the location of error-handling code and the appropriate place to intercept and redirect execution to the error-handling code. Another difference is that these approaches focus on executing past out-of-bounds memory accesses, while Talos, which disables individual functions, can handle a broader set of software faults.

A technique has been proposed to abort the execution of a function when it overruns a memory buffer, as a consequence of malicious inputs, and resume the execution right after the call to the offending function after making a best effort to undo any side-effect caused by the offending function such as changing global variables~\cite{sidiroglou04using}. Their evaluation also indicates that the program can continue run in many situations. The challenge of this work is that many times it is difficult if not impossible to infer what side-effects the partial execution of a function has caused and how to correctly undo them. Talos avoids this problem by not executing any part of a function and simply forcing the function to return an error code to its caller.

\scbf{Finding existing workarounds for failures}
A recently emerging area is searching the configuration space of an application for workarounds for a specific failure.
REFRACT searches for configuration workaround for program failures~\cite{Swanson2014}. Given a model of the configuration space of a program and strategies to avoid failures, REFRACT tries to find a configuration workaround that can avoid the failures caused by malicious inputs, by repeatedly replaying inputs that trigger the failures to the program using different sample configurations. For 6 of 7 Firefox bugs, it successfully found configuration workarounds. Unlike Talos, REFRACT relies on the existing configuration space of a program to workaround a vulnerability. However, our findings show that configurations often do not provide sufficient coverage to workaround most security vulnerabilities. Talos avoids this limitation by instrumenting a program with SWRRs, which are designed specifically to protect the program from being exploited.

\scbf{Automatic patch generation}  A very different approach to solving the vulnerability problem is to try and ensure that patches are always available by automatically generating them.  

ClearView learns invariants of a program in a training phase via dynamic analysis. Once the program is deployed, it monitors for failures and identifies the invariants that are correlated with the failures. If a vulnerability is discovered, it uses the extracted invariants and runtime feedback to generate patches that can be applied to the binaries of the vulnerable programs~\cite{Perkins2009}.

A large body of work also examines the automatic generation of source code patches using vulnerability reports~\cite{Weimer2009, Wei2010, Goues2012-genprog, Goues2012-study, Nguyen2013, Shahriar2012}. 
Based on genetic programming, GenProg takes the source code of a program and a set of test cases as input to construct patches for bugs in programs~\cite{Goues2012-genprog}. A further study on software repair using GenProg achieves a wide range of success rates between 5\% to 100\% for a set of programs. Combining symbolic execution and constraint solving, SemFix synthesizes patches using program synthesis~\cite{Nguyen2013} and achieves an average success rate around 51\%. Their results indicate that creating patches for bugs still largely requires manual work from developers.

On top of fixing vulnerabilities, a recent work aims to misinform attackers about whether an exploit works or not by transforming a regular patch into a honey-patch, which adds additional logic to redirect malicious inputs to a vulnerable version of a program, so that the exploit targeting the patched vulnerability appears to be successful to attackers~\cite{Araujo2014}.

Talos is largely complementary to work on automatic patch generation.  Automatic patch generation faces many of the same difficulties of inferring and maintaining correctness invariants for programs.  The difference is that Talos does not aim to preserve all existing functionality, while automatic patch generation does.

\scbf{Characterizing software vulnerabilities} 
Since software vulnerabilities are such a major source of security threats, there has been a lot of work on characterizing and understanding software vulnerabilities to determine indicators that might predict their presence~\cite{Alhazmi2007, Shin2011, Neuhaus2007}. They measured characteristics such as vulnerability density, defect density, vulnerability discovery rate, structural complexity of code, code churns, and developer activities on code; and built models based on the relationships between them.  We mention these as they served as an inspiration for us to study configuration workarounds to mitigate such vulnerabilities.

A recent study on the life cycle of software releases~\cite{Clark2014} indicates that the rapid-release methodology used by Mozilla Firefox does not increase the ratio of vulnerabilities in the code, somewhat contrary to the popular belief that frequent code changes result in less secure software. Another study has measured characteristics such as evolution of vulnerabilities over the years, impacts of vulnerabilities, and access required for exploits over vulnerabilities and their implications on software design, development, deployment, and management~\cite{Shahzad2012}.

\section{Conclusion}\label{sec:conclusion}
We describe the design and implementation of Talos, a system that enables safe and precise SWRRs to protect software vulnerabilities from being exploited by attackers. Our main conclusion is that SWRRs are a rapid, secure, and low-cost solution to enable applications to continue to be used until a patch becomes available.  To arrive at this conclusion we test 320 SWRRs in five real world applications and find that the majority of them are unobtrusive and that 75.1\% of potential vulnerabilities can be disabled by an SWRR.  This indicates that SWRRs can be effective in 2.1$\times$ more vulnerabilities than traditional configuration workarounds.  We also reproduce 11 vulnerabilities and their exploits and try them on the applications with and without SWRRs instrumented by Talos.  We find that in all 11 cases, the security of the application is upheld and that in 8 cases, the applications retains either all or most of its functionality (with the exception of the vulnerable code).  

We view Talos as a first step towards addressing the pre-patch vulnerability window.  Given its simple implementation and conservative assumptions, we find these results encouraging.  We believe the best avenue for improving the effectiveness of SWRRs is improving the identification of error-handling code or other safe code paths that SWRRs can redirect execution to, which will give SWRRs better basic coverage and thus also increase their effective coverage.  


\section*{Acknowledgements}

We thank our shepherd Gang Tan and the anonymous reviewers for their constructive comments. We also thank Ashvin Goel, Ding Yuan, Michelle Wong, Sukwon Oh for their helpful suggestions and feedback. Financial support for this work was provided in part by funds from a Canada Research Chair and an NSERC Discovery Grant.

\IEEEtriggeratref{62} 
\bibliographystyle{IEEEtranS}
\bibliography{bibfile}

\end{document}